\documentclass[journal=ancac3,manuscript=article]{achemso}
\usepackage[T1]{fontenc}
\usepackage{amsmath}
\usepackage{graphicx}
\usepackage{xcolor}
\usepackage{hyperref}

\author{P\'eter Udvarhelyi}
\affiliation{Department of Chemistry and Biochemistry, University of California, Los Angeles,  CA 90095, USA}
\email{udvarhelyi@ucla.edu}
\author{Prineha Narang}
\affiliation{Division of Physical Sciences, College of Letters and Science, University of California, Los Angeles,  CA 90095, USA}
\alsoaffiliation{Department of Electrical and Computer Engineering, University of California, Los Angeles,  CA 90095, USA}

\title{Design for telecom-wavelength quantum emitters in silicon based on alkali-metal-saturated vacancy complexes}


\begin{document}

\begin{tocentry}
    \centering
    \includegraphics[scale=0.25]{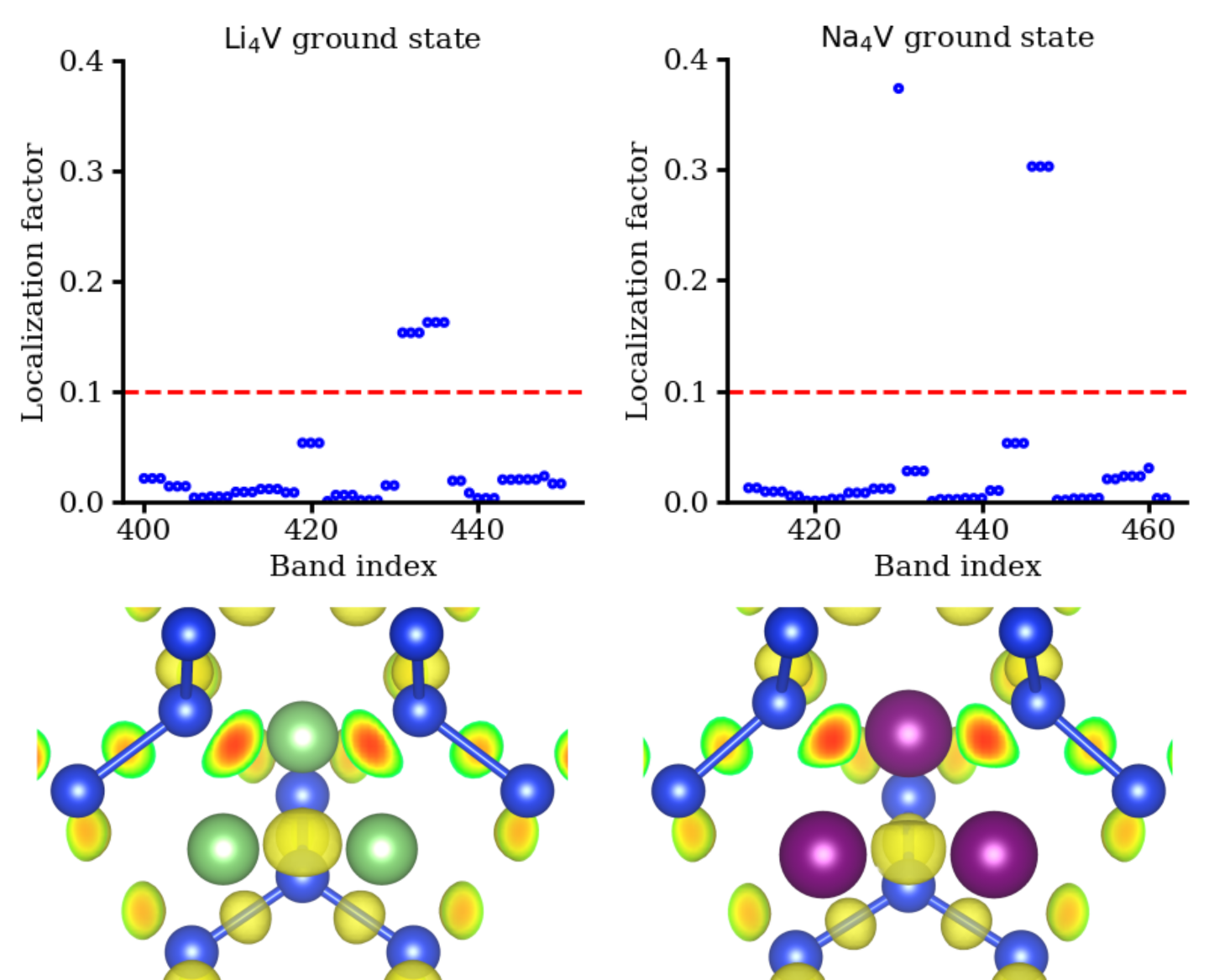}
\end{tocentry}

\begin{abstract}
    Defect emitters in silicon are promising contenders as building blocks of solid-state quantum repeaters and sensor networks. Here we investigate a family of possible isoelectronic emitter defect complexes from a design standpoint. We show that the identification of key physical effects on quantum defect state localization can guide the search for telecom wavelength emitters. We demonstrate this by performing first-principles calculations on the Q center, predicting its charged sodium variants possessing ideal emission wavelength near the lowest-loss telecom bands and ground state spin for possible spin-photon interface and nanoscale spin sensor applications yet to be explored in experiments.
\end{abstract}

Telecom-wavelength single photon sources are key components in quantum cryptography and quantum networks established over large distances~\cite{OBrien_2009}. These take advantage of the currently available optical fiber infrastructure of the internet to transmit quantum information in the form of flying qubits. However, only a relatively narrow emission frequency range allows for ideal attenuation properties in this network. One of the most promising realizations of an on-demand deterministic single photon source is a solid-state defect emitter. Recently, silicon emerged as a versatile material to host several different quantum defect structures emitting in the telecom range, combined with the advantages of a technologically mature CMOS-compatible platform.
Various defect emitters in silicon were thoroughly investigated in experiments decades ago, however, the recent advancements in defect creation~\cite{MacQuarrie_2021, Hollenbach_2022, Baron_2022_apl, Zhiyenbayev_2023} and single center optical measurements~\cite{Redjem_2020, Durand_2021, Higginbottom2022, Komza_2022, Prabhu_2023} renewed the experimental and theoretical interest in these quantum centers~\cite{Beaufils_2018, Bergeron_2020, Ivanov_2022, Redjem_2023, Redjem_2023_cm, Liu_2023, Xiong_2023, Jhuria_2023, Higginbottom_2023}. Several impurity atom complexes were recently identified using first-principles calculations, e.g. di-carbon defects as origins of G~\cite{Udvarhelyi_2021, Deak_2023}, and T~\cite{Dhaliah_2022} lines, the carbon-oxygen defect of C line~\cite{Udvarhelyi_2022}, and the W-center consisting of three self-interstitial atoms~\cite{Baron_2022}. This large diversity of telecom emitters can be mainly attributed to the silicon bandgap lying slightly above the telecom range, favoring pseudo-donor emitters with bound exciton states.
Emitter applications in silicon are dominated by interstitial-related defects, in contrast to the numerous vacancy-related emitter defects in wide bandgap semiconductor hosts. In the following, we focus on the combination of the common vacancy defect structure with isoelectronic bound exciton (IBE) emitter properties. This combination is achieved in a pseudo-donor-acceptor complex, the alkali-metal-saturated vacancy defect family. 

The most common damage center, the vacancy defect can be viewed as a hyperdeep acceptor owing to its unsaturated dangling bonds. One of the first dopants introduced in the silicon crystal was alkali metals for their donor properties~\cite{Korol_1988, Lu_2012}. Among these, Li and Na can be expected to interact with radiation damage defects owing to their relatively fast diffusion in the crystal~\cite{Canham_1988, Fuller_1954}. It was suggested by calculations that Li ions are attracted to the negative charge environment around a vacancy defect forming aggregations~\cite{Tarnow_1992, Morris_2013}. The aggregate containing four Li atoms in a tetrahedral ($\text{T}_{\text{d}}$) configuration was found as the most stable form of this defect complex ($\text{Li}_4\text{V}$)~\cite{Morris_2013}. Its stability was explained by charge transfer from the Li atoms saturating the dangling bonds of the vacancy~\cite{DeLeo_1984, Morris_2013}, forming an isoelectronic neutral defect. It was suggested that this saturation does not completely passivate the vacancy and its $t_2$ defect state should not be repelled far from the bandgap. This was in line with the experimental evidence attributing the neutral singlet IBE state of this defect to the Q zero-phonon line (ZPL) at $1.045~\mathrm{eV}$~\cite{Canham_1980, Canham_1983}.
Additionally, a weak forbidden emission from the triplet IBE state was revealed at $1.05~\mathrm{meV}$ below the Q line in energy and labeled as $\text{Q}_{\text{L}}$. Uniaxial stress splitting of the PL line revealed that the emission is consistent with $\text{C}_{3\text{v}}$ symmetry, attributed to a transition from an $e$ excited state to the $a_1$ ground state level~\cite{Canham_1983, Lightowlers_1984}. The discrepancy between the tetrahedral geometry of the defect and its emission symmetry was explained by symmetry-breaking phonon relaxations in the IBE state~\cite{Davies_1994}. Calculations suggested that only the neutral charge state of the defect is stable~\cite{Morris_2013}. Without the stability of the hole level, the stabilization of the IBE state needs a thorough investigation. With its assigned emission, the defect seems to belong to the recently proposed electrically inactive defect emitter (EIDE) category~\cite{Li_2023}. However, experiments reporting a fast PL decay under heat treatment were explained by charge transition processes~\cite{Kwok_1995}, implying the existence of other stable charge states.

In this paper, we thoroughly investigate the most stable form of Li aggregation in silicon, the $\text{Li}_4\text{V}$ complex in tetrahedral configuration. By applying advanced density functional theory calculations, we aim to identify the origin of the Q emission line in silicon. We model its bonding properties and describe the stability of the IBE. We show that the defect is not an EIDE and its IBE state is connected to the existence of the positively charged defect state and strong electron-phonon coupling possibly allows for the stabilization of its double positive charge state as well. Finally, we search for an ideal emitter analog of the defect based on the unveiled bonding properties of the Q center. We propose its sodium variant, the $\text{Na}_4\text{V}$ defect as an ideal candidate. We connect the stronger localization of the defect orbitals caused by the larger Pauli repulsion and local crystal strain from the larger sodium ions to the position of the defect level. We show that the increased localization opens up a larger stability region for the positively charged states of this defect variant and predict these charged defects as novel telecommunication wavelength emitters with non-zero spin ground states in silicon.

\section{Results and Discussion}
\subsection{Ground state structure}

\begin{figure}[t]
    \centering
    \includegraphics[width=0.8\textwidth]{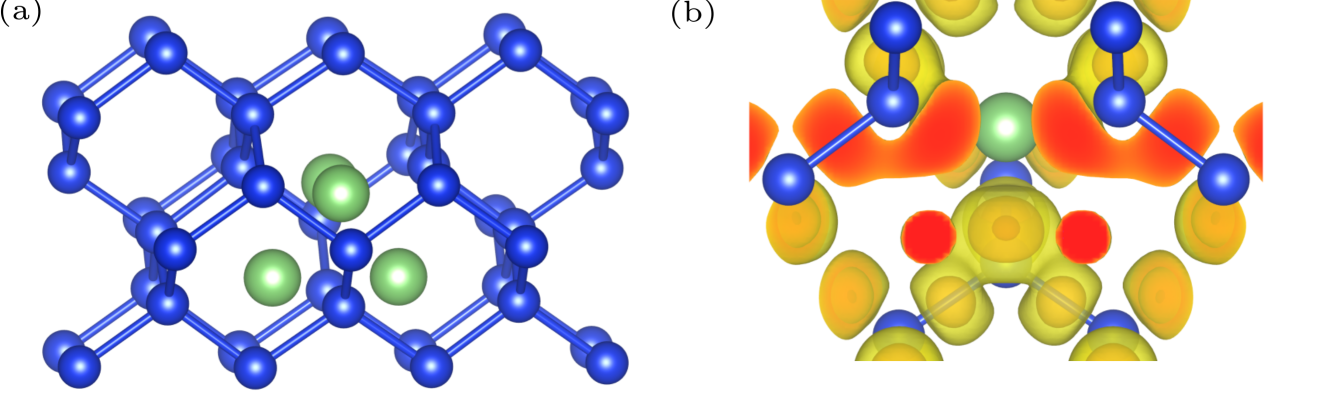}
    \caption{(a) Geometric structure of the neutral $\text{Li}_4\text{V}$ defect in silicon. Li and Si atoms are represented by green and blue balls, respectively. (b) Electron localization function in the neutral ground state of the $\text{Li}_4\text{V}$ defect in silicon. Isovalues of 0.98, 0.95, and 0.85 are drawn in red, orange, and yellow, respectively. As the center of the ELF maxima around the dangling bonds are not lying on the Li-C bond lines, these can be classified as lone-pair orbitals. The $[1\overline{1}0]$ cutoff plane shows the lone pair localization and slight deformation with the in-plane Li atoms removed for clarity.}
    \label{fig:geom}
\end{figure}

The defect model consists of a vacancy site attracting four lithium donors to form a compact aggregate. As the most stable site for the Li impurity in the lattice is at the tetrahedral interstitial position~\cite{Wan_2010}, the energetically most favorable configuration is achieved when the Li atoms occupy the four tetrahedral sites around the vacancy site, contracted to a complex in the center by the interaction with the dangling bonds. The resulting defect structure shows $\text{T}_{\text{d}}$ point symmetry. The Li atoms and the vacancy nearest neighbor Si atoms form intersecting tetrahedrons with Li-Li and Si-Si distances of 2.76~{\AA} and 4.08~{\AA}, respectively. The geometric structure of the defect is shown in Fig.~\ref{fig:geom}(a).

The stability of the defect and the localized levels it introduces into the band structure can be analyzed via the electron localization function (ELF) plotted in Fig.~\ref{fig:geom}(b). It reveals that the Li atoms are ionized in the defect while their donated electron saturates the dangling bonds of the vacancy, in line with previous Mulliken population analysis calculations~\cite{Morris_2013}. However, as the ELF maxima are not located on the lines connecting the Si and Li atoms but rather away from them towards the vacancy site, we conclude that there is no direct bond forming between the two species and the structure is held together by the electrostatic attraction. The saturation of the vacancy dangling bonds forms lone-pair orbitals.

Qualitative analysis of the resulting defect levels can be done in a tight-binding model using symmetry-adapted linear combinations of atomic orbitals (SALCAO). The atomic orbital (AO) basis is the sp3 hybridized Si AOs in $\text{T}_{\text{d}}$ point symmetry. The vacancy introduces four dangling AOs, which span the active space of the tight-binding model. We take a general linear combination of AOs as the initial form of the wavefunction, reading ${\Psi=c_1\phi_1+c_2\phi_2+c_3\phi_3+c_4\phi_4}$. We introduce onsite ($\alpha$) and hopping ($\beta$) integrals between the same and different AOs.

\begin{equation}
    \alpha_{ii}=\int \mathrm{d}^3 r \phi_i(\mathbf{r}) \hat{H}(\mathbf{r}) \phi_i(\mathbf{r}),
\end{equation}
\begin{equation}
    \beta_{ij}=\int \mathrm{d}^3 r \phi_i(\mathbf{r}) \hat{H}(\mathbf{r}) \phi_j(\mathbf{r}),
\end{equation}
where $i\neq j$ are the atomic site indices and the Hamiltonian is written in terms of atomic and perturbing potential $\hat{H}(\mathbf{r})=\hat{H}_{\text{at}}(\mathbf{r})+\Delta U(\mathbf{r})$. All the first neighbor atomic sites are equivalent around the vacancy defect in $\text{T}_{\text{d}}$ point symmetry. Consequently, the indices of the onsite and hopping integrals can be dropped, and hopping connects all the sites. The Schr\"odinger-equation in the AO basis reads

\begin{equation}
    \begin{pmatrix}
        \alpha &\beta &\beta &\beta\\
        \beta &\alpha &\beta &\beta\\
        \beta &\beta &\alpha &\beta\\
        \beta &\beta &\beta &\alpha
    \end{pmatrix}
    \begin{pmatrix}
        c_1\\
        c_2\\
        c_3\\
        c_4
    \end{pmatrix}=\varepsilon
    \begin{pmatrix}
        c_1\\
        c_2\\
        c_3\\
        c_4
    \end{pmatrix},
\end{equation}
where we neglect overlap integrals. The resulting defect levels show a single energy splitting of $4\left|\beta\right|$ between a single and a triple degenerate orbital level. These can be assigned to $a_1$ and $t_2$ irreducible representations in $\text{T}_{\text{d}}$ point symmetry
\begin{align}
    a_1=\left(\phi_1 + \phi_2 + \phi_3 + \phi_4\right)/2,\\
    t_{2x}=\left(\phi_1 - \phi_2 + \phi_3 - \phi_4\right)/2,\\
    t_{2y}=\left(\phi_1 + \phi_2 - \phi_3 - \phi_4\right)/2,\\
    t_{2z}=\left(\phi_1 - \phi_2 - \phi_3 + \phi_4\right)/2.
\end{align}
As $\beta$ is expected to be negative, the totally symmetric $a_1$ defect orbital, corresponding to a bonding configuration, lies lower in energy than the antibonding $t_2$ orbitals. The fully saturated vacancy defect has a closed-shell singlet $^{1}A_1$ ground state which can be accurately described using DFT methods. Even excitation to the conduction band minimum (CBM) and high-spin charged states of the defect are expected to show weak correlation effects. We conclude this qualitative analysis that the level position of the highest occupied defect orbital in the system is determined by the hopping integral which is strongly affected by the localization of the defect orbitals.

Maximum delocalization is energetically preferred for lone-pair orbitals in contrast to the localized dangling bonds. Their resulting defect level is repelled out from the band gap into the valance band. However, owing to the very compact structure of the defect, the ELF of the lone pairs show deformation away from the Li ions. This deformation is commonly observed for the anions in ionic bonds as a result of the Pauli repulsion. Furthermore, this deformation effect from the nearby Li ions makes the lone pair relatively strongly localized comparable to a defect orbital [compare bonding and lone pair localization in Fig.~\ref{fig:geom}(b)]. As we discussed above, the increased localization can be directly connected to the position of the defect level inside the band gap. As the strongly localized lone pair is an intermediate case between the delocalized lone pair and the localized dangling bond, we can expect its level position to lie between the isolated neutral vacancy level position and the lone-pair level inside the valance band near the valance band maximum (VBM), e.g., at a shallow acceptor level position. Our calculations confirm that this is the case for the $\text{Li}_4\text{V}$ defect, where a fully occupied $t_2$ defect level emerges $0.08~\mathrm{eV}$ above the pristine VBM (see left panel in Fig.~\ref{fig:level}). This level originates from the saturation of the in-gap $t_2$ level of the vacancy. The VBM is strongly resonant with the defect level in our calculation leading to its renormalization from the pristine case.

Dissociation of the neutral $\text{Li}_4\text{V}$ defect is modeled by relocating one Li atom from the optimized structure to the nearest neighbor tetrahedral interstitial position and further relaxing the $\text{Li}_{\text{i}}+\text{Li}_3\text{V}$ structure in the neutral spin-polarized singlet state of the whole system. This gives us a lower estimate of $1.24~\mathrm{eV}$ for the dissociation energy. The dissociation is endothermic proving the stability of the initial structure. Upon dissociation, the remaining $\text{Li}_3\text{V}$ aggregate retains its structure close to the original tetrahedral one with the missing Li site. This suggests that the aggregate is formed via one-by-one Li occupation of the tetrahedral sites neighboring the vacancy during the diffusion of Li atoms in the lattice.

Besides its neutral charge state, our calculations show that the defect is stable in a positive charge state with $+/0$ charge transition level at $0.185~\mathrm{eV}$ above VBM. In the ionized state, the defect is $T\times t_2$ Jahn-Teller (JT) unstable leading to a symmetry breaking to $\text{C}_{3\text{v}}$ point symmetry and stabilization energy of $0.08~\mathrm{eV}$. The $t_2^{(3)}$ electronic configuration is modified to $e^{(2)}a^{(0)}$ in the minority spin channel upon ionization. We give a detailed analysis of this effect in the description of the excited state of the defect. Its $2+/+$ charge transition level is only slightly above the VBM after a charge correction energy of $0.3~\mathrm{eV}$ is applied. The ground state of the $2+$ defect is a spin triplet. The removed electron from the degenerate $e$ level of the positive charge state leads to further JT effect of $E\times e$ symmetry. The resulting point symmetry is $\text{C}_{1\text{h}}$ in the $2+$ charge state with a stabilization energy of $0.045~\mathrm{eV}$. The calculated final value of the $2+/+$ charge transition level is only $0.007~\mathrm{eV}$ from which the stability of the charge state cannot be determined within the estimated accuracy of $0.1~\mathrm{eV}$ of our method. However, the spin density of both the doublet and triplet states of the two charged defects are localized in our calculations, in contrast to previous calculations conducted with PBE functional~\cite{Morris_2013}. We attribute this to the better calculation accuracy of the defect level positions using the HSE06 hybrid functional and to the JT stabilization not reported in previous calculations. Our results for at least one or possibly two charged defect states are in line with the experimentally observed fast decay of the Q line photoluminescence after annealing~\cite{Kwok_1995}. The findings can be explained by a large change in the concentration of different acceptor and donor defects inside the silicon sample. An increased substitutional lithium density in the sample, which acts as a shallow acceptor, could pin the Fermi level near the VBM allowing for thermodynamically stable positive charge states of the Q center. Moreover, they reported a series of new photoluminescence lines around the Q line when above bandgap excitation was used which can drive the defect charge state out of the thermal equilibrium related to the Fermi level position. Some of these new lines were stable regarding thermal annealing and proposed to possibly originate from charged Q centers. Our calculation results confirming the existence of at least one charge state of the Q center with spin opens new possibilities for the utilization of the Q center.

\subsection{Optical properties}

\begin{figure}[t]
    \centering
    \includegraphics[width=0.7\textwidth]{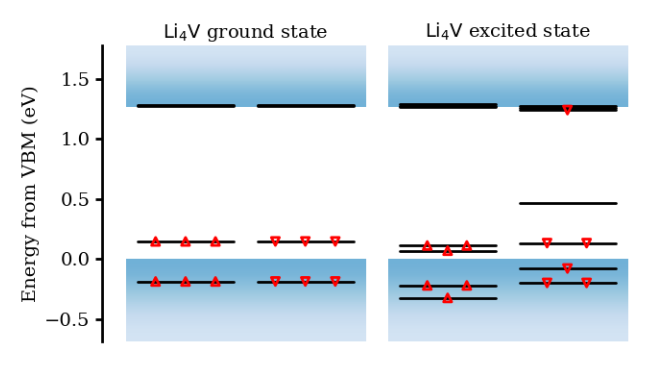}
    \caption{Calculated Kohn-Sham level structure of the $\text{Li}_4\text{V}$ defect in silicon. The $t_2$ orbital of the $\text{T}_{\text{d}}$ defect ground state is resonant with the $t_2$ VBM leading to its renormalization. The excited state is achieved by promoting an electron from the completely filled defect orbital to the CBM. This electron is bound to the hole state, which emerges deep inside the bandgap, forming an isoelectronic bound exciton.}
    \label{fig:level}
\end{figure}

As the defect in its neutral ground state is a closed shell singlet and there are no unoccupied defect states in the bandgap, the lowest energy excited state is reached by promoting an electron from the $t_2$ vacancy orbital to the conduction band minimum (CBM). Consequently, the hole state is localized to the defect and emerges deep inside the band gap, while the electron remains mainly delocalized only slightly perturbing the CBM level (see right panel in Fig.~\ref{fig:level}). The electrostatic interaction however binds this promoted electron to the localized hole, creating a long-living bound exciton. We estimate the exciton binding energy in our Kohn-Sham DFT calculation as the difference between the vertical excitation energy in the $\Delta$SCF calculation and the HOMO-LUMO energy gap, resulting in a bound state with $E_{\text{b}}=0.032~\mathrm{eV}$. We determine the optical lifetime from TDDFT calculation using PBE functional. The obtained $9.3~\mathrm{\mu s}$ is in excellent agreement with the $9.5~\mathrm{\mu s}$ experimental lifetime of the Q center~\cite{Lightowlers_1984}, which is still relatively short compared to other excitonic emitters in silicon~\cite{Davies_1989}.

The strongly localized hole level of the excited state is three-fold degenerate, leading to occupational instability against symmetry-breaking vibrational modes. This is the consequence of strong electron-phonon coupling to the $e$ and $t_2$ modes leading to a descent in symmetry from $\text{T}_{\text{d}}$ to $\text{D}_{2\text{d}}$ and $\text{C}_{3\text{v}}$, respectively. This is formulated in the $T\times (e+t_2)$ Jahn-Teller problem~\cite{Bersuker_2006}. Our excited state relaxation calculations with restricted symmetry (see in parenthesis) result in JT distortion energies of $E_T(\text{C}_{3\text{v}})=79~\mathrm{meV}$, $E_E(\text{D}_{2\text{d}})=5~\mathrm{meV}$ and $E_E(\text{C}_{2\text{v}})=12~\mathrm{meV}$. As $E_T\gg E_E$, we can focus on the dominant $t_2$ single phonon-mode problem of $T\times t_2$. The resulting distortion to $\text{C}_{3\text{v}}$ point symmetry in the excited state can explain the trigonal response of the measured photoluminescence line splitting to uniaxial strain~\cite{Canham_1983}. The relaxed geometries are compared in Table.~\ref{tab:geom}. We note that the IBE excited state is in strong analogy to the ionized charge state, as its properties are mainly determined by the localized hole level. Indeed, the positive charge state of the defect shows the same symmetry of JT instability as described above with the same stabilization energy. The calculated excited state electronic configuration is analogously $e^{(2)}a^{(0)}e_{\text{CBM}}^{(1)}$, in line with the $A$ ground to $E$ excited state transition obtained in uniaxial stress measurements of the PL line~\cite{Canham_1983}.

\begin{table}[t]
    \caption{Geometry comparison between the ground and excited states in high symmetry and after Jahn-Teller distortion. The distances between first neighbor Li atoms and vacancy first neighbor Si atoms are given in {\AA} units. (1) and (2) label basal and axial distances in the corresponding trigonal geometries, respectively.}
    \label{tab:geom}
    \begin{tabular}{lcccc}
        state & Li-Li (1) & Li-Li (2) & Si-Si (1) & Si-Si (2) \\\hline
        ground $\text{T}_{\text{d}}$ & \multicolumn{2}{c}{2.761} & \multicolumn{2}{c}{4.080} \\
        excited $\text{T}_{\text{d}}$ & \multicolumn{2}{c}{2.797} & \multicolumn{2}{c}{4.117} \\
        excited $\text{C}_{3\text{v}}$ & 2.882 & 2.770 & 4.106 & 4.154 
    \end{tabular}
\end{table}

We calculate ZPL energy of $0.99~\mathrm{eV}$ from $\Delta$SCF connecting the excited state in $\text{C}_{3\text{v}}$ point symmetry to the ground state in $\text{T}_{\text{d}}$ point symmetry. This is in good agreement with the experimental Q line energy within the usual $0.1~\mathrm{eV}$ accuracy of such calculations before finite-size corrections. The usual underestimation of the uncorrected ZPL stems from the larger electrostatic potential binding the electron to the hole in the supercell calculation due to confinement effects on the large radius Rydberg state of the electron. More accurate estimation of the ZPL energy requires size scaling correction, which is computationally too expensive at the current level of methods. Furthermore, we calculate the triplet exciton state energy $2~\mathrm{meV}$ below the Q line, corresponding to the experimental $Q_L$ line energy separation of $1~\mathrm{meV}$. Based on the calculated ZPL energy and the symmetry of the optical transition, we unambiguously identify the origin of the Q line in silicon as the $\text{Li}_4\text{V}$ defect.

\subsection{Defect design based on the Q center structure}

The in-depth identification of the Q center provided us insight into the elaborate connection of bonding properties to defect state localization and optical properties. Based on this deeper understanding, we aim to further investigate the possible Q center analog structures from a defect design standpoint. We build our theory on the physical properties described above to engineer the defect charge state stability and level position inside the band gap. These are crucial for enabling a structure with a ground state spin for quantum sensing and tuning its emission wavelength to the favorable telecommunication wavelength bands for minimal signal loss in quantum communication over great distances.

Firstly, we limit our search to the elements with the smallest electronegativity possible, i.e., alkali metals. This is necessary to retain the non-bonding structure of the defect which would be completely passivated otherwise. Next, we consider elements with larger atomic radius to force the localization of the defect orbital even further by the Pauli repulsion. This effect is further strengthened by the larger strain imposed on the vacancy site by the larger metal atoms, leading to an outward relaxation of the nearest neighbor silicon atoms. This relaxation distorts the Si-Si bonding even further from tetrahedral towards a more planar configuration than it would result from the strain field of the vacancy. This again leads to increased localization of the defect orbital owing to the minimization of the lone-pair overlap with bonding orbitals, which are now bent closer. With increased defect orbital localization, we can expect the corresponding defect level to shift deeper inside the bandgap. This effect decreases the ZPL energy compared to the Q center, which lies just above the telecom range, and simultaneously opens up the possibility for thermodynamically stable acceptor levels for spin ground states in the defects. Finally, we consider elements light enough to diffuse in the crystal relatively fast, leading to a high yield of aggregation at the vacancy site after thermal annealing. Based on the above considerations, we select the sodium-saturated vacancy defect ($\text{Na}_4\text{V}$) as an ideal emitter alternative for the Q center.

\begin{figure}[t]
    \centering
    \includegraphics[width=0.75\textwidth]{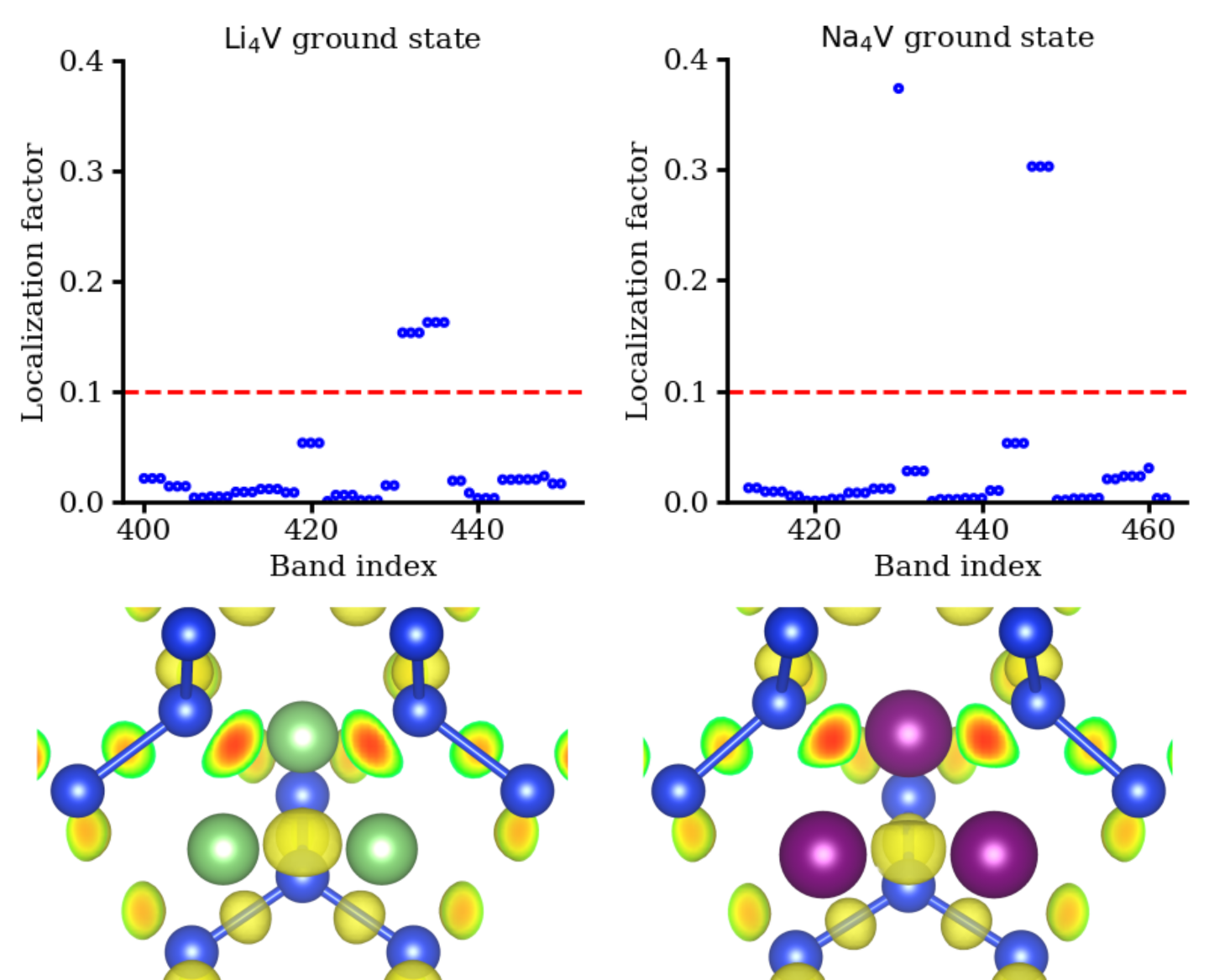}
    \caption{Comparison of defect level localization in the neutral ground state of the Li and Na saturated vacancy defects. Kohn-Sham level localization is assigned above the localization threshold arbitrarily set to 0.1, where the localization factor is calculated in a sphere of 5 a.u. radius around the vacancy site. Below each diagram, the corresponding ELF is plotted at 0.95 isovalue. Li and Na atoms are represented by green and purple balls, respectively. The $[1\overline{1}0]$ cutoff plane shows the colored contour plots of higher isovalues. The larger deformation of the lone pair orbitals for the $\text{Na}_4\text{V}$ defect is in line with the calculated larger localization of the defect Kohn-Sham levels.}
    \label{fig:loc}
\end{figure}

As the microscopic theory behind the physical properties of the $\text{Na}_4\text{V}$ defect is highly analogous to the $\text{Li}_4\text{V}$, we refer to the discussion above and focus only on the differences in the comparison of the two structures. The defect geometry is described by larger Na-Na and vacancy nearest neighbor Si-Si distances of 2.98~{\AA} and 4.51~{\AA}, respectively, which is attributed to the increased strain field of the larger sodium ions. The defect remains stable against dissociation with an even larger endothermic reaction energy of $1.95~\mathrm{eV}$. The localized defect Kohn-Sham level position is shifted deeper inside the bandgap to $0.486~\mathrm{eV}$ above the VBM. To prove that the deeper defect level position is a consequence of increased localization, we calculate the wavefunction localization factor in a 5 a.u. radius around the center of both the defect variants. In Fig.~\ref{fig:loc} we can see a large increase in the localization factor for the Na complex and the absence of strong mixing with the degenerate VBM bands. Our TDDFT calculations result in a much shorter optical lifetime of $0.5~\mathrm{\mu s}$. The one order of magnitude difference compared to the $\text{Li}_4\text{V}$ defect can be attributed to the increased localization of the ground state orbitals.
The considerably deeper defect level and a larger JT distortion energy of $0.116~\mathrm{eV}$ in the excited state result in a much smaller ZPL energy of $0.640~\mathrm{eV}$. In the following, we treat finite size effects as a systematic inaccuracy in the calculations and apply a shift according to the underestimation of the calculated ZPL energy of the Q center before any correction is applied. Taking this into consideration, the ZPL energy in the neutral charge state of the Na complex is $0.695~\mathrm{eV}$. This value undershoots the optimal energy range of the telecom bands. However, it serves as a perfect starting point to explore emission in its positively charged states. As ionization of the defect level introduces a splitting between occupied and empty levels and the occupied defect levels lie lower in energy with respect to the VBM compared to the saturated neutral defect level, consequently their bound exciton states in the CBM show larger ZPL energy. We can naturally expect desaturation to introduce unpaired electrons in the system, leading to high-spin ground states. Furthermore, desaturation of the $t_2$ orbital can lead to occupational instabilities, i.e., several charge states are expected to be JT unstable. The JT stabilization energy of the symmetry-breaking ground states further contributes to the increase of the ZPL energy. We find three stable positive charge states in the system, +, 2+, and 3+ with ground state spins of $S=1/2$, $S=1$, $S=3/2$, and point-symmetries of $\text{C}_{3\text{v}}$, $\text{C}_{1\text{h}}$, and $\text{T}_{\text{d}}$, respectively. The calculated charge transition levels lie at $+/0=0.577~\mathrm{eV}$, ${2+}/+=0.311~\mathrm{eV}$, and ${3+}/{2+}=0.184~\mathrm{eV}$ with respect to the VBM.

Based on the charge stability calculations, two additional BE emissions can originate from the charged $\text{Na}_4\text{V}$ defects. The excited state of the trigonal JT $\text{Na}_4\text{V}^{+}$ defect shows further JT instability ($E\times e$) leading to $\text{C}_{1\text{h}}$ point symmetry. This state is the acceptor-type bound exciton analogy for that of the T center in silicon, which is a donor-type bound exciton~\cite{Bergeron_2020}. As such, it is expected to show reverse roles of the electron and hole interacting with the orbital states. The localized hole state forms a spin triplet with quenched orbital momentum, while the delocalized electron forms two doublet states split from the $J=3/2$ CBM state owing to the low symmetry of the defect in this excited state. As the orbital splitting introduced by this crystal field effect is orientation-dependent with respect to an external magnetic field direction, the observed effective g-factor shows anisotropy in the excited state. This anisotropy is a result of the interplay of two types of JT instability ($T \times t_2$ and $E \times e$) for the $\text{Na}_4\text{V}^{+}$ defect in contrast to the inherently low symmetry structure of the T center. This difference suggests that the dynamic JT effect can thermally quench the magnetic anisotropy in the $\text{Na}_4\text{V}^{+}$ defect. The calculated ZPL energy is $0.719~\mathrm{eV}$. Taking the systematic shift from the Q-line into account, the corresponding shifted ZPL energy is $0.774~\mathrm{eV}$. We predict its experimental ZPL energy lying in the  L telecom band. Its analogous exciton state to the T center makes this defect a promising candidate for telecom spin-photon interface applications~\cite{Higginbottom2022, Dhaliah_2022}.

The second stable charged acceptor bound exciton is formed in the orthorhombic JT, spin-triplet, 2+ charge state of the defect. We can describe the nature of the exciton state similarly to that of the neutral charge state, i.e. the optically allowed transition is from a singlet exciton strongly bound to the additional triplet two-hole system already present at the defect. This picture suggests an isotropic g-factor analog to the Q center. However, in this case, the ground and the optically active excited state both show Zeeman splitting and zero-field splitting in the static JT regime (with axial or lower symmetry). The calculated ZPL and the shifted ZPL are $0.728~\mathrm{eV}$ and $0.783~\mathrm{eV}$, respectively. We can predict its experimental ZPL energy lying in the range from C to S telecom bands. Our calculations also reveal two singlet states at $0.06~\mathrm{eV}$ and $0.34~\mathrm{eV}$ above the triplet ground state.

\begin{table}[t]
    \centering
    \begin{tabular}{lcccc}
	defect &$E_{\text{FC}}~(\mathrm{eV})$&$Q~(\sqrt{\text{amu}}\text{\AA})$&$S$&DWF (\%)\\\hline
$\text{Li}_4\text{V}^{0}$&0.122&1.0072&3.85&2.1\\
$\text{Na}_4\text{V}^{0}$&0.107&1.2741&4.55&1.1\\
$\text{Na}_4\text{V}^{+}$&0.111&1.0726&3.91&2.0\\
$\text{Na}_4\text{V}^{2+}$&0.101&1.0081&3.50&3.0
    \end{tabular}
    \caption{Comparison of phonon interactions in the excitation process for the different defects and their charge states in a single effective mode Franck-Condon approximation. $E_{\text{FC}}$ is defined as the energy difference between the vertical emission and the ZPL, i.e., the ground state phonon relaxation energy. $Q$ is the effective phonon mode connecting the ground and excited state relaxed atomic configurations. $S$ is the Huang-Rhys factor calculated as $S=E_{\text{FC}}/(\hbar \Omega_Q)$, where $\Omega_Q$ is the effective phonon frequency. The Debye-Waller factor (DWF) shows the ratio of the ZPL in the total emission.}
    \label{tab:DWF}
\end{table}

Besides the emission wavelength, another key emitter property for quantum communication is the ratio of the coherent part, i.e. the ZPL, in the total emission. It is measured by the Debye-Waller factor (DWF), related to the Huang-Rhys factor $S$ by $\text{DWF}=\exp(-S)$, which shows the electron-phonon coupling strength in the excitation~\cite{Huang_1950}. These parameters can be estimated by calculating the one-dimensional potential energy surface along the vibrational mode $Q$ connecting the excited and ground state atomic configurations. Our results are summarized in Table~\ref{tab:DWF}. For both defects, we obtain considerable relaxation energy and phonon mode length, resulting in a DWF of only a few percent. This value is far from ideal for an emitter defect, but comparable to several other popular quantum defects, e.g. the negatively charged nitrogen-vacancy (NV) center in diamond~\cite{Alkauskas_2014} and the negatively charged silicon vacancy (V1) in 4H-SiC~\cite{Udvarhelyi_2020}. However, the JT symmetry breaking in the charged states seems to mitigate the relaxation effect in the excitation resulting in a small factor of improvement in the coherence. We also note that waveguide and cavity integration of other defect emitters in silicon was already demonstrated showing a large improvement in the strength and coherence of their emission~\cite{Zhu_2020, DeAbreu_2023, Redjem_2023, Johnston_2024, Gritsch_2023, Saggio_2023, Lefaucher_2023, Lefaucher_2024, Islam_2024}.

With the calculated +/0 charge transition level positioned at the middle of the bandgap, the positive charge state can be thermodynamically stable in semi-insulator or slightly p-type silicon. These conditions could be feasible with focused Na ion beam implantation defect generation that could maximize the aggregation yield and minimize the interstitial Na concentration in the sample, pinning the Fermi level in the lower-gap region by the $\text{Na}_4\text{V}$ acceptor defect concentration. The stabilization of all the positive charge states could also be feasible after thermal annealing at 230$^\circ$ which is predicted to have a large effect on the charge stability of the Q center~\cite{Kwok_1995}. Furthermore, the commonly used above bandgap excitation (532 nm laser) creates a large number of charge carriers in the sample leading to a non-equilibrium charge environment around the defect which can lead to its observable telecom emission even in highly n-doped samples. Recently, electrical stabilization of color centers was demonstrated in a silicon-on-insulator (SOI) diode device~\cite{Day_2024}, which could enable precise control over the charge state of the defect.

\begin{figure}[t]
    \centering
    \includegraphics[width=0.7\textwidth]{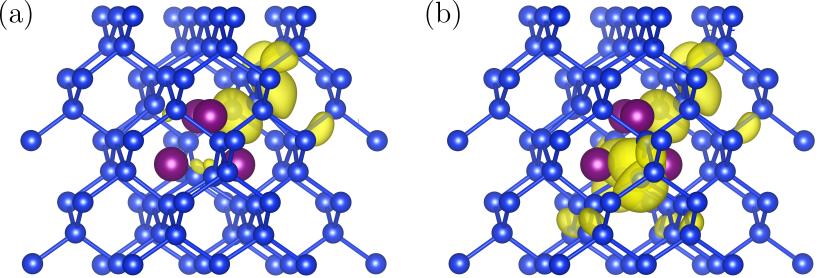}
    \caption{Spin density (yellow) localization defined as the difference between the majority and minority charge density in the spin-polarized calculation. (a) The spin density in the $\text{Na}_4\text{V}^{+}$ defect is localized on the axial Si atom in $\text{C}_{3\text{v}}$ symmetry. The single dangling bond gives rise to a spin doublet. (b) The spin density in the $\text{Na}_4\text{V}^{2+}$ defect is localized on three Si atoms in $\text{C}_{1\text{h}}$ symmetry, resulting in a triplet ground state.}
    \label{fig:spin}
\end{figure}

Finally, we characterize the spin interactions in the spinful charged states of the $\text{Na}_4\text{V}$ defect for spin-photon interface and spin sensor applications. Zero-field splitting (ZFS) is the splitting of magnetic levels ($m_S$) in high-spin states ($S\geq 1$) in axial point symmetry (or lower) without an external magnetic field applied. It originates from the dipolar electron spin interaction, which is extremely sensitive to perturbations in the lattice, thus it is used for nanoscale sensing of various fields in the vicinity of the defect. It might enable spin-selective intersystem crossing transitions to lower spin states creating a spin-photon interface. If this interface can be optically pumped to polarize a specific magnetic level and the manipulation of the spin state creates an optical contrast through the interface, optically detected magnetic resonance (ODMR) is feasible in the defect. For the triplet ground state of the $\text{Na}_4\text{V}^{2+}$ defect, we calculate axial and rhombic ZFS parameters of $D=327~\mathrm{MHz}$ and $E=34~\mathrm{MHz}$, respectively. These findings can be qualitatively explained by examining the spin density plotted in Fig.~\ref{fig:spin}. In Fig.~\ref{fig:spin} (a), we can see that ionization desaturates one of the four lone pair orbitals, restoring a dangling bond and creating a spin doublet radical in the system. This results in a spontaneous symmetry breaking to [111] axial symmetry in the crystal as described above. Removing a further electron restores three dangling bonds, as it is plotted in Fig.~\ref{fig:spin} (b). These dangling bond lobes are directed towards the vacancy site, which explains the rather small $D$ parameter as opposed to a planar dangling bond configuration that creates a large $D$ parameter such as in the case of the NV center in diamond. As the origin of the axial symmetry breaking is the rather weak JT effect, the spin density is highly isotropic with a pseudo rotation axis pointing in the direction of the fourth saturated Si atom. Indeed, our ZFS calculations result in the spin quantization axis of $[-0.568, -0.568, 0.596]$. We note that removing a further electron from the system fully restores the four-dangling-bond configuration of the vacancy in tetrahedral symmetry with a quartet spin ground state. This is the highest stable ionized state in the system.
We also calculate the hyperfine interaction in the $\text{Na}_4\text{V}^{+}$ and $\text{Na}_4\text{V}^{2+}$ defects for the vacancy-nearest-neighbor $^{23}\text{Na}$ and $^{29}\text{Si}$ nuclear isotopes. The results are summarized in Table~\ref{tab:hyperfine}. We obtain large axial hyperfine parameters on the nuclei where the spin density is localized according to Fig.~\ref{fig:spin}. The 3 planar equivalent Na nuclei are promising candidates for nuclear spin memory applications in the axially symmetric $+$ charge state, while the only saturated Si nearest neighbor is likewise promising in the $2+$ charge state.

\begin{table}[t]
    \caption{Eigenvalues of the hyperfine interaction in the $+$ and $2+$ charge ground states of the $\text{Na}_4\text{V}$ defect possessing doublet and triplet spins, respectively. Only the nearest neighbor nuclei of the vacancy site are listed, where symmetrically equivalent $^{23}\text{Na}$ and $^{29}\text{Si}$ nuclei are grouped.}
    \label{tab:hyperfine}
    \begin{tabular}{cccc}
    nucleus     & $A_{xx}~(\mathrm{MHz})$ & $A_{yy}~(\mathrm{MHz})$ & $A_{zz}~(\mathrm{MHz})$ \\\hline
    \multicolumn{4}{c}{$\text{Na}_4\text{V}^{+}$}\\
     1 Na &   16.305542	&16.305542	&18.809742\\
     3 Na &  -1.545248	&-1.832048	&5.765952\\
     1 Si &  -4.823545	&-4.823545	&-125.441745\\
     3 Si &  -1.609119	&-1.580319	&-13.767919\\\hline
     \multicolumn{4}{c}{$\text{Na}_4\text{V}^{2+}$}\\
     1 Na &  4.2536325   &6.2752325	&8.2426825\\
     1 Na &-1.470978    &-1.472828		&-4.962778\\
     2 Na &3.7660625    &5.9553625	&7.7111625\\
     1 Si &-1.091922    &-2.116222		&-53.610422\\
     1 Si &2.5132035    &2.7784535	&3.4853035\\
     2 Si &-2.1759935   &-3.1418435	&-49.2603935\\\hline
    \end{tabular}
\end{table}

\subsection{Strain effect on the optical and spin properties}
Crystal strain is the most common external perturbation on solid-state quantum defects. The precise knowledge of its coupling strength to the magneto-optical properties of the hosted quantum defect is a key ingredient in experiments and applications. On one hand, strong coupling to strain can have a deteriorating effect in identical emitter applications. On the other hand, it can be harnessed for tuning the magneto-optical properties or for precise sensing of the uniform or nanoscale strain environment. We extract these coupling strengths from our first-principles calculation by applying a series of uniform hydrostatic strain magnitudes to the supercell model of the defects. We constrain the lattice parameters to the corresponding strained crystal values, however, we let the local structure of the defect relax under the applied strain. This method takes into account the local variations of the elastic parameters in the vicinity of the defect. The coupling strength parameters are determined by the slope of the response in the linear perturbation theory regime. Our calculations show that this approximation holds well for 1\% strain magnitude. We attribute positive and negative values of strain to expansion and compression, respectively.

We calculate the coupling strength of the ZPL emission to hydrostatic strain for $\text{Li}_4\text{V}^{0}$ and $\text{Na}_4\text{V}^{0}$ defects to be $(1.43\pm0.03)~\mathrm{eV/strain}$ and $(1.21\pm0.01)~\mathrm{eV/strain}$, respectively. As the realistic externally applied strain regime is around 1\%, the tuning of the optical properties is feasible within the 10 meV order of magnitude. Thus, the strain tuning of the Q line to the telecom range is not feasible. These results demonstrate the advantage of the defect design approach compared to tuning the emitter properties via externally applied fields, as the achievable variations in the local electrostatic and strain fields are orders of magnitudes larger for the former.

Furthermore, we calculate the interaction strength of the electron spin with crystal deformation. We obtain $(-260\pm10)~\mathrm{MHz/strain}$ coupling strength between the $D$ parameter and hydrostatic strain. Using the bulk modulus of $97.8~\mathrm{GPa/strain}$ for silicon~\cite{Hopcroft_2010}, the spin-stress coupling strength is $(-2.7\pm0.1)~\mathrm{MHz/GPa}$. This value compares well to that reported for the NV center in diamond~\cite{Barson_2017, Udvarhelyi_2018}, suggesting that the $\text{Na}_4\text{V}^{2+}$ emitter defect is a promising alternative for sensing applications in a silicon platform.

\section{Conclusion}

In this paper, we show that a dramatic change can be achieved in the optical and spin properties of a defect complex, which could not be feasible by external manipulation, by applying a careful defect design guided by a thorough defect identification procedure. To demonstrate this we identify the bonding, charge, and optical properties of the well-known Q center in silicon originating from the $\text{Li}_4\text{V}$ defect complex. We explain the appearance of a shallow in-gap defect level in this system owing to the localized lone pairs on the saturated vacancy-neighbor dangling bonds. We identify two key effects with a large influence on the defect level position inside the band gap, which is crucial for both the bound exciton emission wavelength and the charge state stability of the defect. Namely, Pauli repulsion and local crystal strain leading to deformation of the lone pair orbitals has a large impact on orbital localization and consequently on defect level position. Based on these considerations, we investigate the $\text{Na}_4\text{V}$ defect variant of the Q center structure for ideal emission wavelength. Although the resulting ZPL energy lies just below the long wavelength threshold of the telecom range, the strong Jahn Teller effect in its single and double positive charge states pushes their emission wavelength into the telecom L-band. We show that the sodium variant of the Q center shows a large improvement in the emission strength as well while retaining similar optical coherence. Moreover, its charge states possess non-zero spin ground states crucial for quantum memory and nanoscale sensing applications. We propose the $\text{Na}_{4}\text{V}^{+}$ defect state as a promising candidate for quantum repeater applications, and the $\text{Na}_{4}\text{V}^{2+}$ defect state as a spin triplet strain sensor based on the comparable stress coupling strength of its ZFS to that of the NV center in diamond. Furthermore, the recently demonstrated electrical control of telecom emitters in silicon~\cite{Day_2024} could enable selective functionality of the $\text{Na}_{4}\text{V}$ defect platform based on the precise control of its charge state.

Our work demonstrates an alternate route in the quest for finding ideal emitter defects besides the popular high-throughput search by identifying the key physical effects in defect design and providing a modified defect complex as a promising candidate for experimental investigations.

\section{Methods}
We carried out density functional theory (DFT) calculations using the HSE06 functional~\cite{HSE06} in the plane-wave-based Quantum Espresso code~\cite{QE1, QE2, QE3}. Plane-wave and charge density cutoffs of 80 and 320 Ry were used, respectively. Core electrons were treated using Optimized Norm-Conserving Vanderbilt (ONCV) pseudopotentials~\cite{Hamman_2013, Schlipf_2015}. The atomic positions were relaxed in a 216-atom supercell to $0.001~\mathrm{Ry}/\mathrm{a.u.}$ force threshold. All calculations were performed in the $\Gamma$-point. Excited states were calculated and relaxed using the $\Delta$SCF method of fixed occupations~\cite{Gali_2009}. Size convergence tests on the calculated excitation energies were performed in a 512-atom supercell showing good agreement with the results in the smaller cell. Total energies in charged supercells were corrected using the method of Freysoldt {\it et al.}~\cite{Freysoldt_2009}. Wavefunction localization is calculated in the WEST code, with the localization factor defined as follows 
\begin{equation}
    L_n=\int_V \mathrm{d}^3 \mathbf{r} \left|\Psi_{n}^{(\text{KS})}(\mathbf{r})\right|^2,
\end{equation}
where $n$ is the band index, $\Psi_{n}^{(\text{KS})}(\mathbf{r})$ is the Kohn-Sham wavefunction of a specific band, and $V$ is the integration volume containing the defect~\cite{Sheng_2022}. Optical transition dipoles ($\mathbf{\mu}$) were calculated in the ground state geometric configurations using time-dependent density functional theory (TDDFT) within the Tamm-Dancoff approximation and PBE functional~\cite{PBE} in the WEST code~\cite{Jin_2023}. Optical lifetimes ($\tau$) from the transition dipole moments are obtained using the Wigner-Weisskopf theory~\cite{Weisskopf_1997} as
\begin{equation}
    \frac{1}{\tau}=\frac{2(2\pi)^{3}n_r E_{\text{ZPL}}^{3}\left|\mathbf{\mu}\right|^2}{3\varepsilon_0 h^4 c^3},
\end{equation}
where $n_r=3.5$ is the refractive index of the silicon crystal, $E_{\text{ZPL}}$ is the ZPL energy of the optical transition, $\varepsilon_0$ is the vacuum permittivity, $h$ is the Planck constant, and $c$ is the speed of light. Zero-field splitting (ZFS) is approximated by electron spin dipolar interaction for $S>1/2$ axially symmetric systems, described by the Hamiltonian
\begin{equation}
    \hat{H}_{\text{SS}}=D\left(\hat{S}_{z}^2-\frac{S\left(S+1\right)}{3}\right)+E\left(\hat{S}_{x}^2-\hat{S}_{y}^2\right)\text{,}
\end{equation}
where $D$ and $E$ are the axial and rhombic ZFS parameters, respectively. The calculation is performed in the spin-polarized HSE ground state with ONCV potentials using the PyZFS code~\cite{PyZFS}.
The interaction between the electron spin and nearby nuclear spins is described by the hyperfine interaction. Its Hamiltonian after diagonalization reads
\begin{equation}    \hat{H}_{\text{IS}}=A_{xx}\hat{I}_{x}\hat{S}_{x}+A_{yy}\hat{I}_{y}\hat{S}_{y}+A_{zz}\hat{I}_{z}\hat{S}_{z}\text{.}
\end{equation}
Hyperfine parameters ($A$) are calculated using PAW potentials~\cite{Blochl_1994} and PBE functional within the GIPAW module of Quantum Espresso. Visualization of the atomic structures is done in VESTA~\cite{VESTA}.

\section{Data availability}
Calculation data is available from the authors upon reasonable request.

\section{Author contributions}
P. Udvarhelyi conceived the project idea, carried out the calculations, analyzed the data, and wrote the manuscript. P. Narang supervised the project, and acquired funding and computational resources. All authors discussed and edited the manuscript.

\section{Acknowledgment}
The authors acknowledge the support of the National Science Foundation QuSeC-TAQS under Award No. 2326840.
This research used resources of the National Energy Research Scientific Computing Center, a DOE Office of Science User Facility supported by the Office of Science of the U.S. Department of Energy under Contract No. DE-AC02-05CH11231 and Award No. BES-ERCAP0029123.

%\bibliography{bib}

\providecommand{\latin}[1]{#1}
\makeatletter
\providecommand{\doi}
  {\begingroup\let\do\@makeother\dospecials
  \catcode`\{=1 \catcode`\}=2 \doi@aux}
\providecommand{\doi@aux}[1]{\endgroup\texttt{#1}}
\makeatother
\providecommand*\mcitethebibliography{\thebibliography}
\csname @ifundefined\endcsname{endmcitethebibliography}  {\let\endmcitethebibliography\endthebibliography}{}

\end{document}